\documentclass[aip,pop,amsmath,amssymb,
reprint,%
]{revtex4-1}
\pdfoutput=1

   \usepackage[pdftex]{graphicx}
   \usepackage{epstopdf}

\usepackage{dcolumn}
\usepackage{bm}

\begin{document}

\pagestyle{plain}

\title[]{Diamagnetic ``bubble'' equilibria in linear traps}

\author{A.D. Beklemishev}
\email{bekl@bk.ru}
 \altaffiliation[Also at ]{Novosibirsk State University.}
\affiliation{Budker Institute of Nuclear Physics SB RAS, Novosibirsk, Russia}


\begin{abstract}
The plasma equilibrium in a linear trap at $\beta\approx 1$ (or above the mirror-instability threshold) under the topology-conservation constraint evolves into a kind of diamagnetic ``bubble''. This can take two forms: either the plasma body greatly expands in radius while containing the same magnetic flux, or, if the plasma radius is limited, the plasma distribution across flux-tubes changes, so that the same cross-section contains a greatly reduced flux. If the magnetic field of the trap is quasi-uniform around its minimum, the bubble can be made roughly cylindrical, with radius much larger than the radius of the corresponding vacuum flux-tube, and with non-paraxial ends. Then the effective mirror ratio of the diamagnetic trap becomes very large, but the cross-field transport increases.  The confinement time can be found from solution of the system of equilibrium and transport equations and is shown to be $\tau_E\approx\sqrt{\tau_\parallel\tau_\perp}$. If the cross-field confinement is not too degraded by turbulence, this estimate in principle allows construction of relatively compact fusion reactors with lengths in the range of a few tens of meters. In many ways the described here diamagnetic confinement and the corresponding reactor parameters are similar to those claimed by the FRCs. 

\end{abstract}



\maketitle

\section{\label{sec:Introduction}INTRODUCTION}
Though tokamaks are now undisputed leaders of the fusion program, their design has intrinsic limitations that deny possibility to work with advanced fuels such as $d-d$ or aneutronic $p-^{\rm 11}$B. While further progress of tokamaks is pinned to slow construction of gigantic ITER, the small linear traps and field-reversed configurations (FRCs) demonstrate steady and rapid progress beyond what was just 10 years ago thought to be their absolute ceiling.\cite{GDT1k,C2U} The road to fusion for these ``alternative'' plasma traps is certainly still very long and rocky, but based on current theory and scalings they can offer an attractive vision of the more compact reactor with high energy density and relative engineering simplicity. This paper describes a novel concept of efficient high-beta plasma confinement that is half-way between the FRC and the linear gas-dynamic trap (GDT). It certainly improves on the standard GDT confinement by offering a greatly reduced reactor size at the cost of a more tricky MHD stabilization. The energy confinement in the diamagnetic ``bubbles'' will be probably worse than that in FRCs, but initial estimates suggest that a less stringent maintenance of equilibrium and stability will be required.

The diamagnetic ``bubble'' equilibrium is a new mode of operation of gas-dynamic linear traps. There is currently just one trap of this type in operation, the GDT in Novosibirsk.\cite{Ivanov} It was originally constructed for concept exploration of a fusion neutron source for materials testing. The idea was to use beam-beam fusion in the population of fast ``sloshing'' ions, injected and adiabatically confined in an axisymmetric mirror, while stability of the anisotropic beam plasma was ensured by the relatively cold core component. Confinement of the core plasma due to its high density and low temperature would be in the collisional gas-dynamic regime,
when the outflow of the ions is limited by the nozzle effect of the mirror throats, while the electron heat flux along open field lines is limited by the ambipolar electrostatic potential. The theory of the axial gas-dynamic confinement predicts the total loss of about $8kT_e$ per an ion-electron pair escaping to the end walls, and is experimentally confirmed.\cite{Ryutovaxial,GDTaxial} The GDT reached and surpassed its design goals. Its success is mainly a consequence of the simple and robust design. Collisional plasma tends to be more stable, while additional turbulent scattering, if present, cannot further increase the already maximum possible gas-dynamic outflow rate. The record plasma parameters in GDT are now $\beta=60\%, T_e\lesssim 1$keV, at $n_e\sim 10^{19}m^{-3}$, with mean ion energy of $8$keV.\cite{GDT1k} These parameters are essentially sufficient for construction of the fusion neutron source for materials science. The successful GDT design may even be used for a neutron driver of nuclear waste burner or a hybrid reactor.\cite{Anikeev} However, its reliance on the beam-beam fusion limits the GDT fusion efficiency to $Q_{DT}\lesssim 1$.

Nevertheless, the fusion prospects of gas-dynamic traps were considered.\cite{R,Ryutovreaktor} Discarding the beam-beam fusion, but retaining the gas-dynamic confinement of the core plasma leads to the linear scaling of fusion efficiency with the trap length, $Q_{DT}\propto L$, so that the gas-dynamic reactor is theoretically possible. Unfortunately, the size of such a reactor is too large ($L> 5$km) and its power is enormous as well. To alleviate this deficiency it is necessary to improve the axial confinement of ions. For reduction of the plasma outflow even the early papers suggested to use multiple mirror plugs.\cite{MM,R} This idea was reborn after the efficiency of the mutiple-mirror confinement was shown to be at least two orders of magnitude better than expected at target plasma densities\cite{GOL-3}. Indeed, the axial flux reduction by a series of $N$ mirrors has its maximum of $\sim N$ times, when the ion scattering length is equal to the distance between mirrors. In presence of self-consistent plasma turbulence this seems to be satisfied even at densities much lower than predicted via binary collisions. A combination of the gas-dynamic core with multiple-mirror end plugs was dubbed GDMT\cite{GDMT} and promises quadratic scaling with length: $Q_{DT}\propto L^2$. If the scaling is true, the GDMT reactor with DT fuel may be just $300$m long with plasma radius of about $10$cm, and the power is acceptable. However, it is disputable that the self-consistent turbulence that provides additional scattering in mirror cells of the GOL-3 trap will be the same under reactor conditions with $N\sim 100$. An advanced concept with active helical mirror plugs that allow better flux reduction will be tested in SMOLA device.\cite{helic}

While improvement of efficiency of mirror plugs may be in theory sufficient for construction of a gas-dynamic fusion reactor, its parameters at this stage are not attractive enough to justify the risk of a new investment into lagging-behind-tokamaks linear traps. The main drawback is in the geometry: while it is easy to construct an axially symmetric tube-like reactor, it has to be long and thin. The reasons are as follows: the fusion power is proportional to the plasma volume and squared density, $n$ , while the lost power is proportional to the plasma cross-section in mirror throats and density. The plasma occupies a magnetic flux-tube, so that cross-sections of the mirror throats and of the active zone are related as the ratio of the magnetic fields, i.e., the mirror ratio $R=B_{m}/B_0$. As a result, $Q_{DT}\propto nRL$. Now the maximum density as well as the mirror ratio are related to the maximum attainable confining magnetic field. Indeed, in the paraxial approximation the equilibrium is limited by $\beta$: $n\propto \beta B_v^2$, where $B_v$ is the confining (vacuum) field in the active zone, while the magnetic field within the plasma is reduced as $B_0=B_v\sqrt{1-\beta}$. Thus
\begin{equation}
Q_{DT}\propto LB_vB_m\frac{\beta}{\sqrt{1-\beta}}.\label{1}
\end{equation}
It follows that both the mirror field, $B_m$, and the confining field, $B_v$, should be chosen as high as technically possible, while the plasma radius can be made small as long as transverse losses stay less than axial. This last requirement is actually determining the length-to-radius ratio of the optimized reactor, $L/a\gg 1$.

\begin{figure}[h]
\includegraphics[width=0.95\columnwidth]{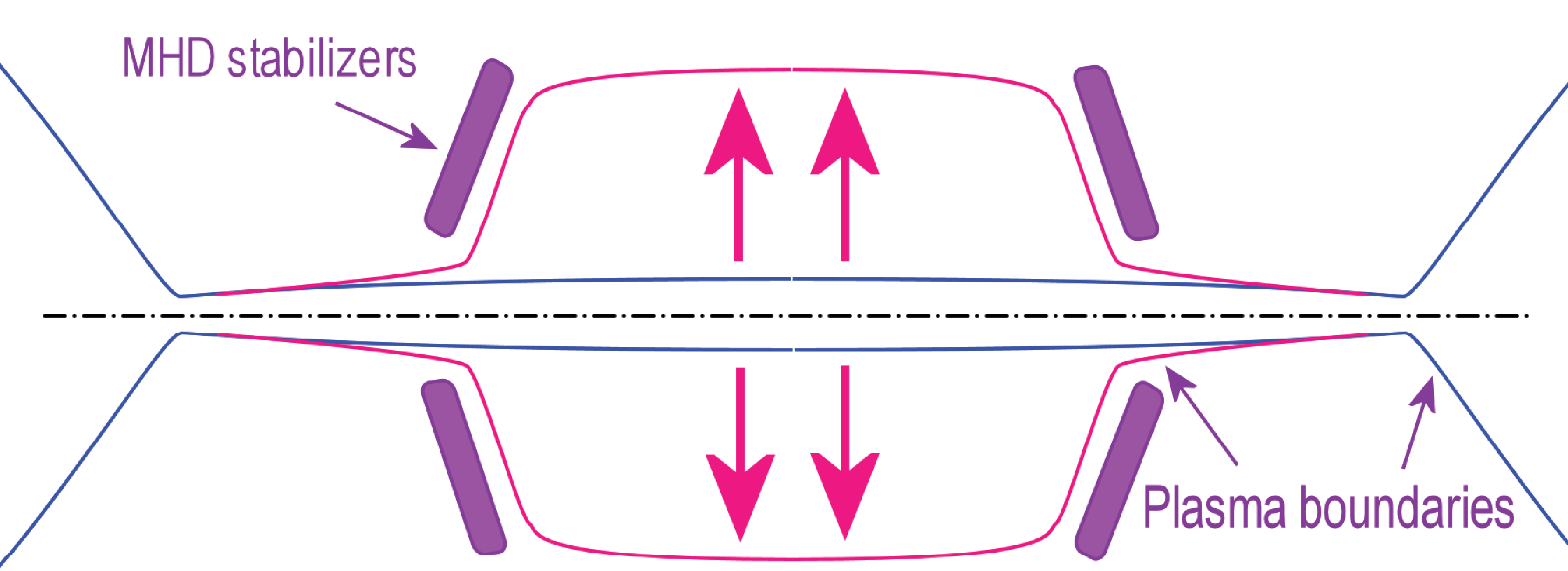}
\caption{Expansion of flux tubes at high $\beta$ leads to corresponding increase in the effective mirror ratio of a linear trap. If there is a quasi-uniform patch of the vacuum field at the bottom of the magnetic well, the resulting ``bubble'' will be roughly cylindrical. The plasma boundary at cylinder ends needs stabilization. \label{puzyr}}
\end{figure}

One can notice a very interesting $\beta$-dependent factor in Eq.(\ref{1}). While $\beta$ is far from unity, say less than $60\%$ as in GDT, the gain in $Q_{DT}$ from $\beta$ is linear. However, as $\beta\rightarrow 1$ the effective mirror ratio of the trap starts to grow rapidly due to diamagnetic radial expansion of the flux-tubes (see Fig.\ref{puzyr}). This effect has been noticed at least 50 years ago,\cite{TW} but at the time most linear traps were adiabatic ones with logarithmic dependence of confinement time on the mirror ratio. Little importance was attached to the fact. 
However, it may be a game-changer for the gas-dynamic confinement due to linear scaling of $Q_{DT}\propto R \propto 1/\sqrt{1-\beta}$. The increase in fusion efficiency due to $\beta$ can be translated into a corresponding decrease in $L$, i.e., a really compact fusion reactor based on a linear trap may become possible.

Before going too optimistic about the $\beta\rightarrow 1$ limit, we should answer a lot of difficult questions. The first is: how large the $1/\sqrt{1-\beta}$-factor can be made in realistic equilibria? Furthermore, even if we reach this stage, it may not be as beneficial as expected. While the magnetic flux is expelled from the plasma core, the transverse transport is also bound to increase up to infinity as the ions become unmagnetized. Will the gain in axial confinement be sufficient to justify the increased radial diffusion? In fact, this diffusion is the same as the diffusion of the confining magnetic field into the diamagnetic ``bubble'', and the radial structure of the equilibrium will be defined by particle sources. The force balance at $\beta\approx 1$ is so delicate, that it is impossible to properly describe the plasma equilibrium state without considering the balance of particle and energy fluxes. Of course, the existence of an equilibrium state does not guarantee that it can be realized in experiment. It should be made stable at least to the ideal MHD modes. This feat is also very tricky at high $\beta$, as the predicted limit of ballooning stability in linear traps may be significantly lower than 1.\cite{Ball1,Ball2,Ball3}

At this point it is time to remember that new ideas are often old ones in disguise, and the history tends to repeat itself. Once upon a time (about 40 years ago), there was a period, when one of the mainstream brands of plasma devices for fusion was the linear $\theta$-pinch. The magnetic configuration of this confinement scheme is the same as that of linear traps. The linear $\theta$-pinches are actually direct predecessors of the multiple-mirror traps and FRCs that were born in attempts to tackle high axial losses. In terms of the $\theta$-pinch community these are particular types of ``end-stoppering'' schemes. The recently invented helical mirror plug\cite{helic_first} would qualify as a special sort of ``peristaltic mirror''. Due to typically high plasma density, above $10^{22}m^{-3}$, the axial outflow from $\theta$-pinches was in the gas-dynamic regime (far from the mirror throats). In this sense the GDT can also be considered a hybrid of the $\theta$-pinch for the core plasma with the adiabatic mirror for hot ions. The most important point is that $\theta$-pinches were inherently  high-$\beta$ devices, and the $\beta\rightarrow 1$ limit with its $\sqrt{1-\beta}$ factor in axial confinement was noticed\cite{TW} and explored in detail theoretically\cite{TW,W,Morse,Freidberg,Steinhauer,Tajima,Taj2}, experimentally,\cite{E1,E2} and by advanced numerical simulation.\cite{Tajima,Taj2} The general conclusion, which one can derive from reading papers of the $\theta$-pinch era on the subject, is that if one tries really hard and is lucky, one can at best expect a factor of 2 to 5 improvement in plasma confinement due to the high-$\beta$ effect. Any improvement is nice, but this is a bit disappointing.

So what is new in linear traps as compared to $\theta$-pinches to warrant revival of this old and forgotten idea? The main difference is the transient-decay regime of operation of a typical $\theta$-pinch versus ``stationary'' operation regime of a linear trap. The actual numbers are not as much different, like 100 microseconds for a $\theta$-pinch, vs. 5 milliseconds for the current GDT, but ``trap''-based reactors are definitely stationary in contrast to ``pinch''-based pulsed reactors. The high-temperature phase of a discharge in a linear $\theta$-pinch continued just as long as its axially expanding plasma column was separated from the end-walls by vacuum. Once the electrical contact between them was established, two bad things happened: the electron temperature plummeted, due to recirculation with the cold plasma at the wall and the high parallel electron conductivity, and the plasma rotation appeared with an often unstable ``wobbling'', due to radial currents driven by  ambipolar potentials. The initial high-temperature phase was stable, but only if there were no mirrors at the ends. A stable operation with mirrors was also possible, but only if the mirrors were turned on {\em after} the electrical contact with the end-wall, to ensure the line-tying stabilization. The most important advances made by the GDT with respect to this picture are the abilities to thermally insulate the internal electrons from the end-wall by electrostatic barriers in expanders, to control the radial distribution of plasma potential and rotation, to routinely work in stable regimes with unfavorable-on-average curvature and high mirror ratios, and to maintain the approximate particle and energy balance.

The transient mode of operation of $\theta$-pinches, and the instability, caused by the presence of end mirrors, made their imprints on the way the high-$\beta$ effect was treated. The primary attention was given to transient phenomena, such as rarefaction waves, in uniform vacuum fields (without external end mirrors). Another consequence was the treatment of the radial profile of plasma pressure as an initially generated function that would decay rather than evolve toward some stationary state. In particular, the scaling of the life-time $\tau\propto\sqrt{\tau_\perp\tau_\parallel}$ was obtained by Steinhauer,\cite{Steinhauer} but only in the context of flux enhancement by rarefaction waves due to transverse diffusion in thin plasma columns with sharp radial gradients. Thus it is time to look at the high-$\beta$ effect on confinement in linear systems from the new angle of stationary equilibria that is relevant for modern traps. 

 This paper describes the high-$\beta$ equilibrium in linear traps, the way to reach it, and possible important benefits to use this mode of confinement for the design of fusion reactors. The plasma behavior in the $\beta\approx 1$ limit is very different from that in more common regimes and lacks established description. Thus, the results here are all obtained in zero-order approximation and should definitely be refined by future work. In particular, the influence of the parallel plasma flow on the form of the equilibrium and on its stability as well as most kinetic effects are left for the future.
 
 In section II the ``bubble'' formation in an anisotropic plasma with the increase of pressure beyond the mirror-instability threshold is described. The cylindrical (rather than spherical) ``bubble'' with non-paraxial ends can be formed in traps with quasi-uniform field near its minimum. In section III the requirements on heating power for successful transit into the ``bubble'' regime are discussed. Section IV is devoted to description of the saturated stationary ``bubble'' by solution of the equilibrium and particle-balance equations. The results lead to the estimates of the energy confinement time and possible reactor parameters. In section V one possible approach to stabilization of the diamagnetic ``bubble'' is formulated. In conclusion the results are summarized and concept-exploration experiments are discussed.

\section{HIGH-$\beta$ PARAXIAL EQUILIBRIA IN LINEAR TRAPS}
A lot of facts is already known about equilibria in linear traps.\cite{Post} Our aims here are limited and very specific: to study the high-$\beta$ limit of equilibrium in long and thin axially symmetric traps. It appears that this particular limit typically results in significant growth of the initially thin plasma in radius. That expansion is axially localized around the minimum of the vacuum magnetic field.

The transverse pressure of plasma in a linear axisymmetric trap can be approximated as
\begin{equation}
p_{\perp }=\int f\left( \varepsilon ,\mu ,P_{\varphi }\right) mv_{\perp
}^{2}d^{3}v\approx p_{\perp }\left( \psi ,B\right), \label{pperp}
\end{equation}
where $\psi$ is the flux function labeling magnetic surfaces, and $B$ is the local magnetic field strength.
Then the equation of paraxial equilibrium looks like
\begin{equation}
B_{v}^{2}=B^{2}+8\pi p_{\perp }\left( \psi ,B\right) , \label{equil}
\end{equation}
where $B_{v}(\vec{r})$ is the confining vacuum magnetic field. 

 Let's assume that we are interested in a subset of solutions that models a sequence of equilibria in a given trap with growing pressure. One typical case is when the distribution function is produced by inclined neutral beam injection (NBI) into the  minimum of the magnetic field, like in GDT. Since the field minimum at finite pressure, $B_0$, itself depends on the distribution function via the equilibrium, the function $p_{\perp }\left( \psi ,B\right) $ will be different for different pressures. Going back to Eq.(\ref{pperp}), one can see that in a strong field what really matters is the pitch angle that varies along field lines according to the local field strength. Then the adequate representation of the sequence of equilibria with growing pressure but constant injection angle is $p_\perp=p_\perp(\psi,R)$, where $R\left( \psi ,\ell \right)=B/B_0$ is the local mirror ratio of the magnetic field, and the function $p_\perp(R)$ retains its form. Note that the subsequent use of this approximation does not alter the generality of description, since for a given equilibrium $B_0$ is just a constant.
  
Let's divide the equilibrium equation (\ref{equil}) by the square of the minimum vacuum magnetic field on a field line, $B_{v0}^2$, and normalize the pressure and the magnetic field by their values at the field minimum (trap center). We get
\begin{equation}
R_{v}^{2}\left( \psi ,\ell \right) =R^{2}\frac{B_{0}^{2}}{B_{v0}^{2}}+\beta
_{0}P\left( \psi ,R\right) ,
\end{equation}
where $R_{v}\left( \psi ,\ell \right)=B_v/B_{v0}$ is the mirror ratio of the vacuum field, $\beta _{0}\left( \psi \right) =8\pi
p_{\perp }\left( \psi ,1\right) /B_{v0}^{2}$ is the
$\beta $-value at the minimum of the field on a given field line, while $P\left(
\psi ,R\right) =p_{\perp }\left( \psi ,R\right) /p_{\perp
}\left( \psi ,1\right) $ is the normalized profile of pressure along the field line.
Note that this equation should be valid everywhere, including the field minimum ($R=1,R_{v}=1,P=1$), thus
\[
1=\frac{B_{0}^{2}}{B_{v0}^{2}}+\beta _{0}. 
\]
Finally we arrive at
\begin{equation}
R_{v}^{2}\left(\ell \right) =\left( 1-\beta\right) R^{2}+\beta
P\left(R\right), \label{ravn}
\end{equation}
where the common argument $\psi$ and the zero subscript of $\beta$ are omitted to shorten notation.
This equation should be solved in order to find $R\left( \ell\right) $ with restriction $R>1,$ while we are 
particularly interested in the case $1-\beta \ll 1.$

Let's assume that we are dealing with a typical mirror with a monotonically growing field from its center. Then 
the left-hand side grows with $\ell$, $\partial R_v^2/\partial \ell \geq 0$, and to find a solution for all $\ell$ we should have a growing right-hand side too,
\begin{equation}
\frac{\partial R^2}{\partial \ell}\cdot\frac{\partial }{\partial R^2}\left[ \left( 1-\beta \right) R^{2}+\beta
P\left( R\right) \right] \geqslant 0. 
\end{equation}
This solubility condition results in a restriction on the pressure profile along the field line:
\begin{equation}
\frac{\partial P\left( R\right) }{\partial
R^{2}}\geqslant -\frac{1-\beta }{\beta }.  \label{crit}
\end{equation}

In linear traps there are always some areas, where the pressure derivative along the field line is negative. Indeed, the pressure should be higher inside of the trap than in the mirror throats, since otherwise there would be no point in using mirrors. Looking at Eq.(\ref{crit}), one can see that such decreases of pressure are restricted, and at $\beta\rightarrow 1$ they are entirely prohibited by the paraxial equilibrium. What is the reason for this restriction? Does it mean that there are no high-$\beta$ equilibria we are looking for?

\begin{figure}[h]
\includegraphics[width=0.9\columnwidth]{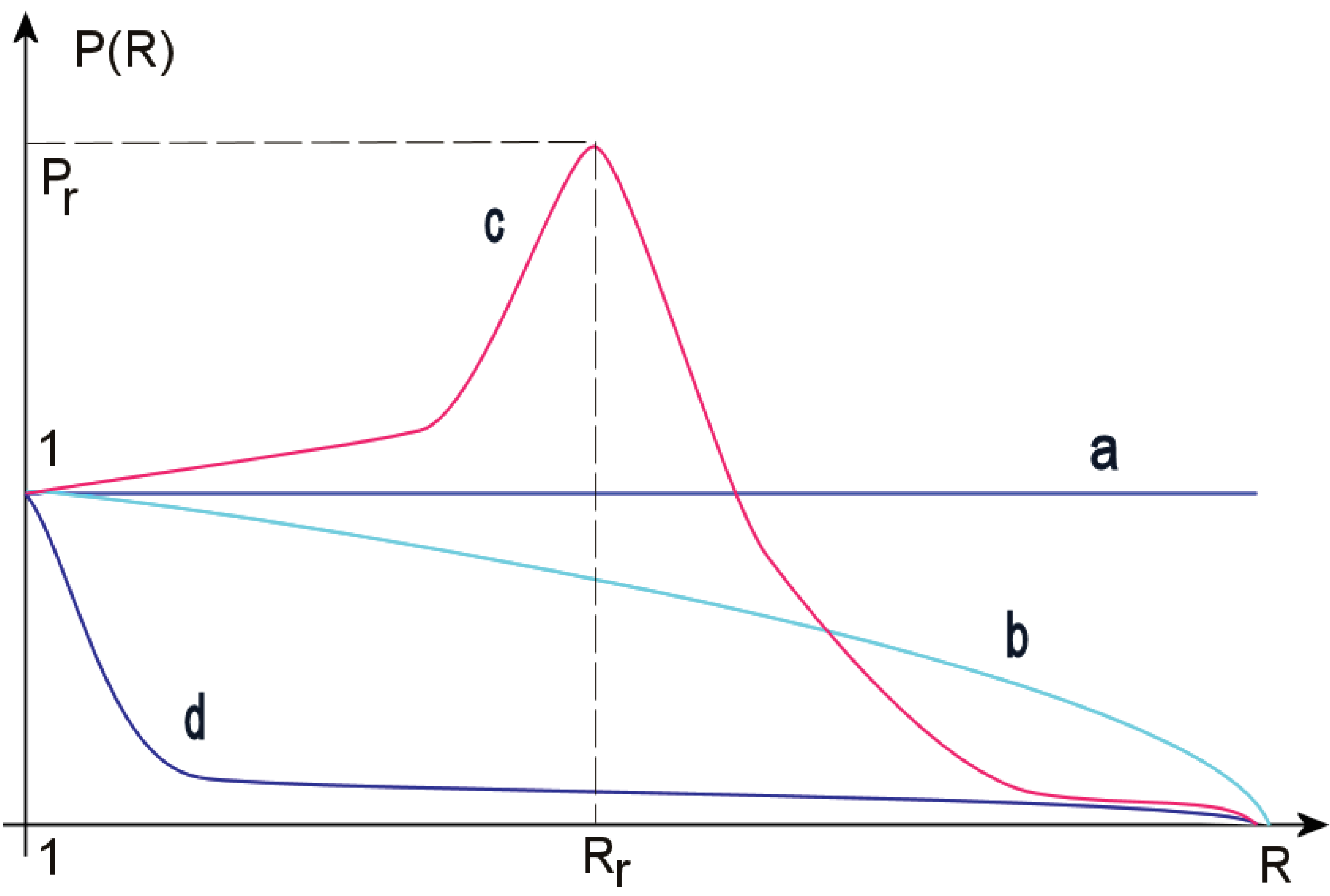}
\caption{Typical forms of the normalized transverse pressure distribution on a field line vs. mirror ratio $R=B/B_0$ for different distribution functions: a) isotropic; b) with a loss cone; c) with inclined injection at $\alpha={\rm arctan}{\sqrt{R_r-1}}$ to the normal; d) with normal injection. \label{anis}}
\end{figure}

The answer to these questions can be based on analysis of Kotelnikov\cite{Kot1,Kot2}. According to his papers the equilibrium solutions
of the paraxial equilibrium equations can be piecewise continuous, while the points of discontinuity can be interpreted as non-paraxial areas (with sharp inclination of field lines to the magnetic axis). At high $\beta$ the function $R\left( \ell \right) $ becomes discontinuous. It is comprised of two (or more) continuous intervals: the internal one, where condition (\ref{crit}) is satisfied  at low $R\sim 1,$ and the external ones, where the same condition is satisfied as well, but at large $R$ only. Note that at $\beta\rightarrow 1$ the acceptable values of the mirror ratio $R$ inside the trap can become arbitrarily large. Since 
$P\left( R\right) >0$, for all reasonable distribution functions
\[
\lim_{R\rightarrow \infty }\frac{\partial P\left( R\right) }{\partial R^{2}}%
=0, 
\]
and thus the piecewise continuous equilibrium solutions exist for all
$\beta <1.$ Let's describe some such solutions for typical types of pressure anisotropy
shown in Fig.\ref{anis}.

\subsection{Quasi-isotropic distribution functions and normal injection}
In these cases the function $P\left( R\right) $ is monotonously decreasing from 1 to 0, while the fully isotropic case corresponds to $P\equiv 1,$ i.e., for all $R$ we have
\[
\frac{\partial P\left( R\right) }{\partial R^{2}}\leqslant 0. 
\]

In the isotropic case solutions of Eq.(\ref{ravn}) are continuous for all $\beta<1$:
\begin{equation}
R(\ell)=\sqrt{\frac{R_{v}^{2}\left( \ell \right) -\beta }{1-\beta }}. 
\end{equation}
However, as $\beta\rightarrow 1$ the solutions become localized in the vicinity of the bottom of
the magnetic well on a field line:
$R_{v}^{2}\left( \ell\right)
\sim 1.$ If the solution width $\ell_d$ is defined by $R^{2}(\ell
_{d})= h,$ then
\[
R_{v}^{2}\left( \ell _{d}\right) =\left( 1-\beta \right) h+\beta \rightarrow
1. 
\]
This isotropic solution cannot be directly applied to open traps due to pressure anisotropy there, but, as we shall see,
the feature that the equilibrium tends to collapse toward the bottom of the magnetic well on a field line is quite robust.

In presence of anisotropy, let's define 
\[
\left( \frac{\partial P\left( R\right) }{\partial R^{2}}\right)
_{R^{2}=1}=-\lambda ^{2}. 
\]

\paragraph{If $\lambda ^{2}\neq 0$} (the normal injection case), the equilibrium equation at $R=1$ 
takes the form
\[
R_{v}^{2}\left( \ell \right) =\left( 1-\beta \right) R^{2}+\beta -\lambda
^{2}\beta \left( R^{2}-1\right) . 
\]
If $\beta <1/\left( 1+\lambda ^{2}\right) ,$ 
solution is continuous:
\[
R^{2}(\ell)=1+\frac{R_{v}^{2}\left( \ell \right) -1}{1-\beta \left( 1+\lambda
^{2}\right) }. 
\]
Above this threshold, if $\beta \geqslant 1/\left( 1+\lambda
^{2}\right) ,$ solution acquires discontinuity at $\ell_d$, where $R_{v}(\ell_d)=1,$ i.e.,
at the bottom of the magnetic well. $R(\ell<\ell_d)=1$, while immediately beyond it $R=R_{d},$ which is the mirror ratio of the ``bubble''. It satisfies
\[
1=\left( 1-\beta \right) R_{d}^{2}+\beta P\left( R_{d}\right) . 
\]
In the limit $1-\beta \ll 1,$ $R_d$ tends to infinity:
\[
R_{d}\approx 1/\sqrt{ 1-\beta} \gg 1. 
\]

\paragraph{If $\lambda ^{2}=0$} (the quasi-isotropic case), one can approximate the pressure profile by parabola:
\[
P\left( R\right) \approx 1-\delta ^{2}\left( R^{2}-1\right) ^{2}. 
\]
Then the equilibrium equation becomes
\[
R_{v}^{2}\left( \ell \right) =1+\left( 1-\beta \right) \left( R^{2}-1\right)
-\delta ^{2}\beta \left( R^{2}-1\right) ^{2}. 
\]
This quadratic equation has a positive solution for $R^{2}-1$ if
$
D=\left( 1-\beta \right) ^{2}-4\delta ^{2}\beta \left( R_{v}^{2}\left( \ell
\right) -1\right) \geqslant 0. 
$
While the ``bubble'' solution is not point-like like in the previous case, it is still discontinuous, and
the length to discontinuity is defined by
\[
R_{v}^{2}\left( \ell _{d}\right) =1+\frac{\left( 1-\beta \right) ^{2}}{
4\delta ^{2}\beta }. 
\]
One can see that  $R_v(\ell_d)\rightarrow 1$ with $\beta \rightarrow 1$, i.e., the ``bubble'' branch of solution collapses to the
bottom of the magnetic well as before.

\subsection{Distribution function with sloshing ions}
In this case $P\left( R\right) $ has a local maximum at reflection point of fast ions, $R_{r}$, and then
decreases to zero due to the loss-cone effects. Where $P(R)$ grows, there is always a continuous solution, while 
on the downward slope at sufficient $\beta$ discontinuity can appear. Its position will tend to $R_r$ as $\beta\rightarrow 1$.

Let's denote $P\left(
R_{r}\right) \equiv P_{r},$ and use parabolic approximation of $P(R)$ near the local maximum:
\[
P\left( R\right) \approx P_{r}-\lambda ^{2}\left( R-R_{r}\right) ^{2}. 
\]
Note that this approximation is reasonable only for the purpose of finding position of discontinuity at $\beta\rightarrow 1$ (since it will be close to $R_{r}$.)
The equilibrium equation near $R_r$ looks like
\begin{equation}
R_{v}^{2}\left( \ell \right) =\left( 1-\beta \right) R^{2}+\beta \left(
P_{r}-\lambda ^{2}\left( R-R_{r}\right) ^{2}\right) . \label{slosh}
\end{equation}
The presence of discontinuity can be detected as an absence of solutions in some interval of $R$ due to negative 
discriminant of this quadratic equation.

{If $\beta \geqslant 1/\left( 1+\lambda^{2}\right) ,$} solution of Eq.(\ref{slosh}) becomes discontinuous. Position of discontinuity can be found from condition that the discriminant is zero:
\[
R_{v}^{2}\left( \ell _{d}\right) =\beta P_{r}+\frac{\lambda \beta \left(
1-\beta \right) }{\beta \left( 1+\lambda ^{2}\right) -1}R_{r}^{2}. 
\]
It appears that though the discontinuity shifts toward the bottom of the magnetic well with growing pressure,
it always stays on the downward slope of $P(R)$ beyond the reflection point.
In the limit $\beta \rightarrow 1$ the ``bubble'' branch of solution is still finite-length:
\begin{equation}
R_{v}^{2}\left( \ell _{d}\right) \rightarrow P_{r}>1. \label{pr}
\end{equation}
It is interesting to note that its axial extent depends on the peaking of the pressure profile rather than on the injection angle.

\subsection{Analysis and interpretations}
At first glance the results of the above section look unusable due to appearance of discontinuities and zero-length solutions. But these zero-length singular solutions at the bottom of the magnetic well are not as bad as they look. As noted by Kotelnikov,\cite{Kot1,Kot2} discontinuity of the paraxial solution is just a non-paraxial transition between two branches of the equilibrium. Thus, solution that has a branch with a single point at $R_v=1$ (the ``bubble'' that collapsed to the bottom of the magnetic well at $\beta=1$), can be interpreted as just a roughly-spherical ``bubble'' in the middle of the trap, if $R_v(\ell)$ is parabolic. 

This last condition, describing the function $R_v(\ell)$, is extremely important. It describes the form of the magnetic well of the vacuum field of the trap along field lines. Near the bottom in a typical mirror it is indeed parabolic. However, we can choose this function according to our aims. In particular, we can design a linear trap with a finite-length patch of uniform field at the well bottom. Then 
\begin{equation}
R_v(\ell)\equiv 1,\qquad {\rm for}~~ \ell<\ell_0,
\end{equation}
and the branch of equilibrium that exists only at $R_v=1$ becomes extended into a cylinder of length $\ell_0$. The best thing about this trick is that the ``bubble'' length can thus be prescribed via the form of the vacuum field. The bubble edges at high $\beta$ will coincide with the ends of the uniform-field patch, so that we will be able to place there some equipment for MHD stabilization.

The type of high-$\beta$ equilibrium that appears with sloshing ions formally allows a finite-length ``bubble'' even with a parabolic $R_v$ profile. However, this is not quite so: with growing $\beta$ the magnetic field in the confinement area of beam ions is decreased, while the magnetic field at reflection points becomes non-paraxial. Under such conditions the conservation of the magnetic moment will be very poor. As a result, the pressure anisotropy will relax, so that $P_r\rightarrow 1$. As one can see from Eq.(\ref{pr}), this means that the developed ``bubble'' with inclined injection will be no different from a typical quasi-uniform case, only the path to it will vary.

In the following sections the cylindrical geometry of the diamagnetic ``bubbles'' and the trap design with quasi-uniform patch of the magnetic field will be assumed. The other possible case of a roughly spherical non-paraxial ``bubble'' seems to lack prerequisites for plasma stabilization (to be discussed in Section V). In particular, in the cylindrical case the interchange source term is finite only at the ends of the cylinder, while in the middle uniform patch the plasma is marginally stable. Thus, by increasing length we can add plasma without worsening stability. Furthermore, by placing localized stabilizers directly at the ends of the cylinder it should be theoretically possible to suppress even the edge ballooning modes. The inclined NBI that leads to formation of sloshing ions (like in GDT) seems also to be far from optimal. Indeed, in this case more of the "bubble" will extend over the field area with unfavorable curvature, so that the overall configuration will be less stable. Normal injection may also be bad, as the resulting cylindrical bubble may be unstable to splitting into spherical ones by the mirror instability. Thus, the trap configuration should have a uniform stretch of field in the middle and any NBI should be inclined at about $60..80^o$ to the axis, so that the resulting pressure profile looks like curve b) in Fig.\ref{anis}. 

\section{\label{sec:TR}PATH TO THE ``BUBBLE'' REGIME}

Let's write the energy and particle conservation laws within a flux tube
of length $L$ and cross-section $S$, assuming gas-dynamic axial losses:
\begin{eqnarray}
\frac{d}{dt}\left[ \left( \frac{3}{2}p+\frac{B^{2}}{8\pi }\right) SL\right]
+\left( p+\frac{B^{2}}{8\pi }\right) \frac{d}{dt}\left[ SL\right]=\nonumber \\
 =W-8pv_{m}S_{m},
\end{eqnarray}
\begin{equation}
\frac{d}{dt}\left( nSL\right) =Q-2nv_{m}S_{m}.\label{n}
\end{equation}%
Here $W$ and $Q$ are the source terms, $v_{m}$ and $S_{m}$ are the outflow
velocity and the flux-tube cross-section in the mirror throat, $p$ is the
total (quasi-isotropic) plasma pressure, $n$ is the average ion density. The
energy loss rate through two mirror throats is taken to be proportional to $8T_{e}$ per escaping ion (as in GDT), that translates into $4p/n$ if $T_{e}\approx T_{i}.$ Additional relations can be obtained from the paraxial
equilibrium,
\begin{equation}
8\pi p+B^{2}=B_{v}^{2}=const,
\end{equation}
and the flux conservation, 
\begin{equation}
BS=B_{v}S_{0}=B_{m}S_{m}=const.
\end{equation}

This simplified system is full. It is also obvious that the equation for
density separates from the system if we can ignore the dependence of escape
velocity on temperature. Strictly speaking
$
v_{m}\approx \sqrt{p/nM_{i}}.
$
However, the evolution of density in time may influence $W$ in real
experiments. For the sake of obtaining a simple estimate we shall assume
that the source of particles is configured to keep the plasma temperature
approximately constant, so that $v_{m}$ is constant too.

Then we can obtain a single equation of evolution of the flux-tube
cross-section. Let $Y=S/S_{0}=B_{v}/B,$ then  
\begin{equation}
\beta =1-1/Y^{2},
\end{equation}
and
\begin{equation}
\frac{1}{2}\frac{d}{dt}\left( \beta Y\right) +2\frac{dY}{dt}=\frac{8\pi W}{
S_{0}LB_{v}^{2}}-8\beta \frac{v_{m}S_{m}}{LS_{0}}.
\end{equation}
After substitution
\begin{equation}
\dot{Y}=2\frac{w-\nu \left( 1-1/Y^{2}\right) }{5+Y^{2}},  \label{ydot}
\end{equation}
where $w=8\pi W/S_{0}LB_{v}^{2}$ is the ratio of heating power to initial
magnetic energy within the flux tube, and $\nu =8v_{m}S_{m}/LS_{0}$ is the
inverse gas-dynamic confinement time in the vacuum magnetic configuration.
It is quite obvious that in order to reach the ``bubble''-type equilibrium
with $Y\gg 1$ one should have
\begin{equation}
w/\nu =\frac{\pi W}{v_{m}S_{m}B_{v}^{2}}\gtrsim 1.  \label{wv}
\end{equation}
This condition is independent of $L,$ but is proportional to the density of
heating power per unit cross-section of the initial flux tube and is inversely
proportional to the magnetic energy density. 

Condition (\ref{wv}) describes total power requirements to heat plasma as a
whole. There is a less stringent scenario based on gradual heating of the
thin central flux tube. Assume that we can focus our heating up to a fixed
power density $w_{d}$, so that the total heating power in the flux tube
grows with its cross-section, $w=w_{d}Y$ (as long as this cross-section is
less than some limit). In this case the stationary states of Eq.(\ref{ydot})
satisfy
\begin{equation}
\frac{Y_{s}^{2}-1}{Y_{s}^{3}}=\frac{w_{d}}{\nu }.
\end{equation}
If we gradually increase the ratio $w_{d}/\nu $, we obtain gradually higher $
Y_{s}$ roots, i.e., stationary ``bubble'' cross-sections. But only below some threshold power density. The
maximum of the left-hand side is reached at $Y_{sc}^{2}=3,$ and if $
w_{d}/\nu >2/3\sqrt{3}$, there are no more stationary roots, so that the
flux tube will keep expanding (up to the limit of cross-section). This condition can be
rewritten as
\begin{equation}
w_{d}/\nu =\frac{\pi \left( W/S_{0}\right) }{v_{m}B_{v}^{2}}R_{m}>0.4,
\label{wdv}
\end{equation}
where $W/S_{0}$ is the required heating density, and $R_{m}=S_{0}/S_{m}$ is
the vacuum mirror ratio of the trap.

Condition (\ref{wdv}) describes the required power density of plasma heating
to reach the ``bubble'' state, if the confinement is gas-dynamic. However,
even in GDT the confinement quality is better than that due to the
population of fast ions. The threshold $\beta ,$ that has to be surpassed
(at the bottom of the magnetic well), might be a better indicator: 
\begin{equation}
\beta _{c}=1-1/Y_{sc}^{2}=2/3.
\end{equation}

\section{\label{sec:Co}RADIAL STRUCTURE OF A SATURATED ``BUBBLE''}
Let's try to describe an axisymmetric steady-state equilibrium taking into
account diffusion of the external magnetic field into the \textquotedblleft
bubble\textquotedblright . If the magnetic field is nevertheless stationary,
this means that there is a steady flux of plasma from the inside, i.e., the
radial losses. The radial plasma flux across the magnetic field is governed
by the azimuthal component of the Ohm's law:

\[
E_{\varphi }+\frac{1}{c}\left[ \vec{v}\times \vec{B}\right] _{\varphi }=%
\frac{j_{\varphi }}{\sigma };
\]%
where $\sigma $ is the effective transverse conductivity, $j_{\varphi }$ is
the azimuthal current density, $\vec{v}$ is the flow velocity, and the
azimuthal electric field $E_{\varphi }=0,$ if the magnetic field is
constant. Let's define the flux function $\psi $ according to%
\[
\left[ \vec{B}\times \vec{e}_{\varphi }\right] =-\frac{1}{2\pi r}\nabla \psi
;
\]
then%
\[
\vec{v}_{\perp }=-2\pi rс \frac{j_{\varphi }}{\sigma }\frac{\nabla
\psi }{\left\vert \nabla \psi \right\vert ^{2}}
\]%
is the velocity of plasma slipping through the magnetic field. The flux of
plasma ions across a magnetic surface $\psi =const$ is then%
\begin{equation}
\Phi =\mathbf{-}2\pi с \oint \frac{nr}{\sigma }\frac{j_{\varphi }}{%
\left\vert \nabla \psi \right\vert }dS.  \label{Phi}
\end{equation}%
The balance of the number of ions in a flux-tube $d\psi $ can be written as 
\begin{equation}
\frac{\partial \Phi }{\partial \psi }=\frac{\partial }{\partial \psi }\left(
Q-2nv_{m}S_{m}\right) =Q^{\prime }-2nv_{m}/B_{m},  \label{Phi1}
\end{equation}%
i.e., any divergence of the radial flux is due to the external source and
axial losses within the flux-tube. The axial losses are taken to be
gas-dynamic, as in Eq.(\ref{n}).

The same azimuthal current density that enters into equation for the flux, (%
\ref{Phi}), also enters the MHD equation of the transverse equilibrium%
\begin{equation}
j_{\varphi }=2\pi rc\frac{\partial }{\partial \psi }p_{\perp }\left( \psi
,B\right) ,  \label{jp1}
\end{equation}%
and can be linked to distribution of $\psi $ via Maxwell equation 
\begin{equation}
j_{\varphi }=-\frac{c}{4\pi }r\left( \nabla \cdot \frac{\nabla \psi }{2\pi
r^{2}}\right) .  \label{jp2}
\end{equation}%
Equations (\ref{jp1}),(\ref{jp2}) together form an analog of the
Grad-Shafranov equation for axisymmetric linear traps%
\begin{equation}
\Delta ^{\ast }\psi =-16\pi ^{3}r^{2}\frac{\partial }{\partial \psi }%
p_{\perp }\left( \psi ,B\right) .  \label{GS}
\end{equation}

Strictly speaking, one has to solve the system (\ref{Phi})-(\ref{jp2}) in
real geometry together with some realistic model describing relationship
between the transverse pressure $p_{\perp }\left( \psi ,B\right) $ and the
ion density $n.$ However, as a first approximation, we will formulate and
solve a drastically simplified system. Simplifications are based on
considerations given in the previous sections: 1) We consider a
\textquotedblleft developed bubble\textquotedblright\ in a magnetic field
with a long uniform patch of length $L$. The field geometry will be close to
a straight cylinder, while contribution to transverse flux from its
ends can be neglected. 2) The effective mirror ratio within a
\textquotedblleft developed bubble\textquotedblright\ is expected to be very
large, so that the plasma pressure is almost isotropic and the axial losses
are indeed gas-dynamic. 3) For the sake of simplicity we take the equation
of state to be $p=2nkT$ with $T=const,$ so that $\sigma =const,$ $v_{m}=const
$ as well. 4) The ion source is assumed to be provided by pellet injection
and is located somewhere deep inside the \textquotedblleft
bubble\textquotedblright , so that only its integral rather than radial
distribution matters.

In these approximations $B=\psi ^{\prime }/2\pi r$ and the ion balance
equation (\ref{Phi1}) simplifies to     
\begin{equation}
\left[ \frac{np^{\prime }r}{B^{2}}\right] ^{\prime }-\frac{2\sigma nv_{m}Br}{
c^{2}LB_{m}}=0,
\end{equation}
where prime denotes derivative in radius. The Grad-Shafranov equation also
simplifies and yields the familiar paraxial approximation,
$B^{2}=B_{v}^{2}-8\pi p$,
where $B_{v}$ is the magnetic field outside of the plasma cylinder.

Excluding the magnetic field profile from the ion balance equation, and introducing $
\beta \left( r\right) =8\pi p/B_{v}^{2}$ to replace $p$ and $n,$ we get a
single nonlinear equation%
\begin{equation}
\left[ \frac{\beta \beta ^{\prime }r}{1-\beta }\right] ^{^{\prime }}=\lambda^{-2}r\beta 
\sqrt{1-\beta },  \label{bp1}
\end{equation}
where 
\begin{equation}
\lambda ^{-2}=\frac{16\pi \sigma v_{m}B_{v}}{c^{2}LB_{m}}.  \label{bp2}
\end{equation}
The characteristic radial scale $\lambda$ can be rewritten as 
\[
\lambda =\sqrt{\frac{c^{2}}{4\pi \sigma }\frac{LR_{m}}{4v_{m}}},
\]
and can be interpreted as a skin depth of the magnetic field by the time of
the gas-dynamic outflow from the vacuum field of the trap. It is normally
very small, for example, for $T=300eV,$ $R=35,$ $\mu =2,$ $L=3$ m we find $\lambda \approx 6.5mm.$ It should be even smaller for fusion parameters.  

Equation (\ref{bp1}) can be rewritten as a system
\begin{eqnarray}
y^{\prime } &=&\lambda ^{-1}r\beta \sqrt{1-\beta }, \\
\beta ^{\prime } &=&\lambda ^{-1}y\left( 1-\beta \right) /\beta r.  \nonumber
\end{eqnarray}
The boundary conditions are $\beta \left( \infty \right) =0,$ $y\left(
0\right) =y_{0}<0,$ where $y_{0}$ is the (almost on--axis) source of ions.
Qualitatively, solution for the \textquotedblleft developed
bubble\textquotedblright\ looks as follows: over most of the radius $\beta
\approx 1,$ while the transition layer from $\beta \approx 1$ to $\beta
\approx 0$ (the boundary) has the characteristic radial scale $\lambda \ll r.
$  This structure can be successfully described in the slab approximation,
i.e., we set $r\approx a=const$, $\lambda ^{-1}y/a=f,$ and introduce the
normalized radial coordinate $x=\left( r-a\right) /\lambda $. Now%
\begin{eqnarray}
f_{x}^{\prime } &=&\beta \sqrt{1-\beta },  \label{bp3} \\
\beta _{x}^{\prime } &=&f\left( 1-\beta \right) /\beta ,  \nonumber
\end{eqnarray}
describes the structure of the boundary layer.

The systems (\ref{bp1}),(\ref{bp3}) are nonlinear boundary problems.
Fortunately, (\ref{bp3}) is simplified to the extent that it can be
integrated in quadratures. We first find equation for $f\left( \beta \right) 
$:
\[
fdf=\frac{\beta ^{2}d\beta }{\sqrt{1-\beta }},
\]
that can be integrated with boundary conditions:
\begin{equation}
f^{2}=\frac{2}{15}\left[ 8-\sqrt{1-\beta }\left( 8+4\beta +3\beta
^{2}\right) \right] .  \label{bp4}
\end{equation}
Note the important limiting value 
\begin{equation}
f\left( \beta =1\right) =4/\sqrt{15}\approx 1.03  \label{fb}
\end{equation}
that describes the normalized particle source to keep the given
\textquotedblleft bubble\textquotedblright\ stationary.

\begin{figure}[h]
\includegraphics[width=0.9\columnwidth]{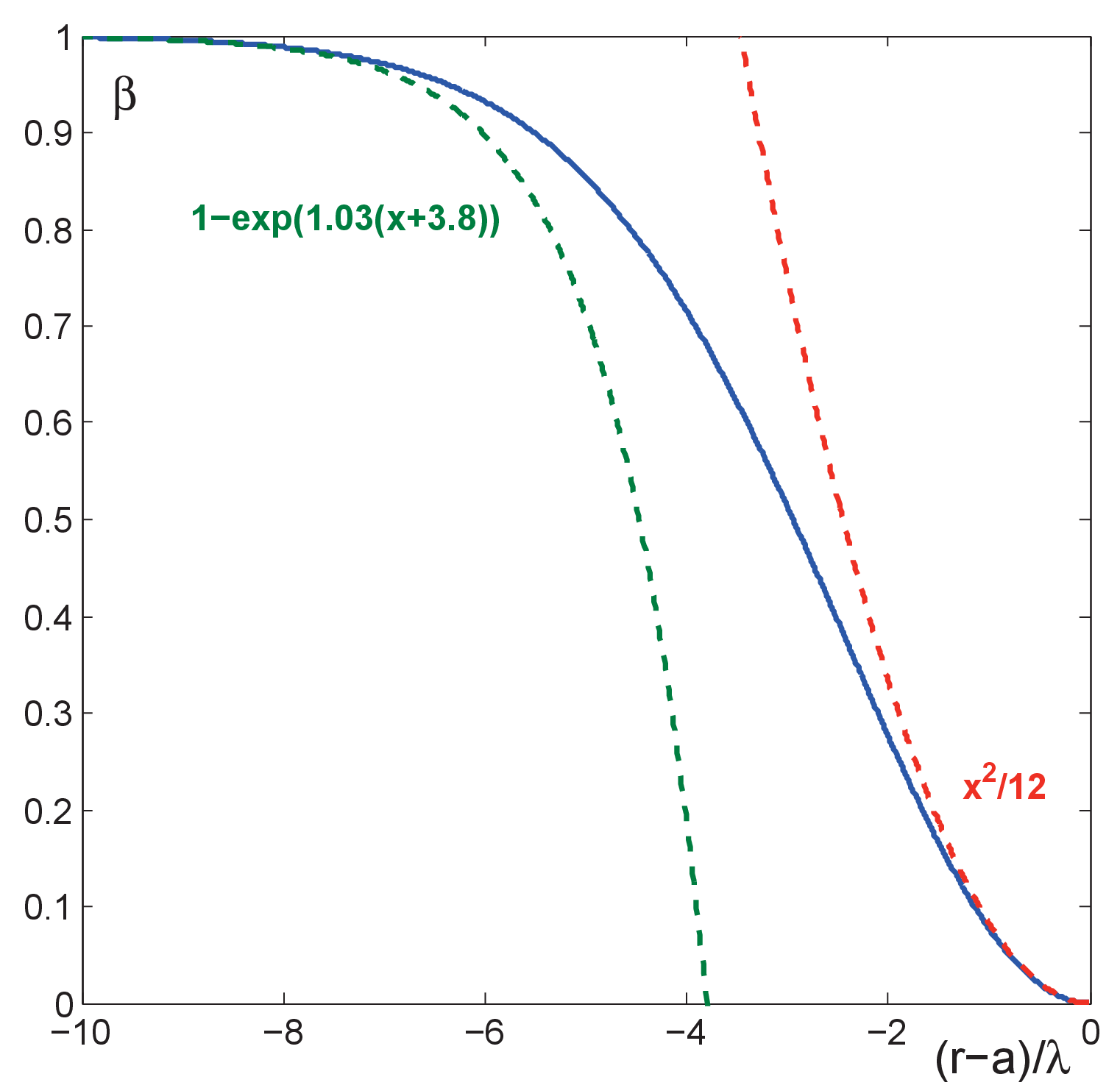}
\caption{The radial structure of the boundary layer of the ``bubble'' in the MHD slab-transport model. There is vacuum ($\beta=0$) beyond the $r=a$ surface, and the low-field interior ($\beta\approx 1$) to the left. \label{sloi}}
\end{figure}

Substituting Eq.(\ref{bp4}) into the second line of Eq.(\ref{bp3}), we get
\begin{equation}
\beta ^{\prime }=-\frac{4}{\sqrt{15}}\frac{1-\beta }{\beta }\sqrt{1-\sqrt{
1-\beta }\left( 1+\frac{\beta}{2}+\frac{3\beta ^{2}}{8}\right) }.
\end{equation}
This equation can only be solved numerically (see Fig.\ref{sloi}), however, the
asymptotics can be found analytically as
\begin{equation}
1-\beta \approx \exp \left[ \frac{4}{\sqrt{15}}\left(\frac{r-a}{\lambda }+3.8
\right) \right] ,\quad \frac{r-a}{\lambda }\rightarrow -\infty ;
\end{equation}
\begin{equation}
\beta \approx \frac{\left( r-a\right) ^{2}}{12\lambda ^{2}},\quad -1<\frac{
r-a}{\lambda }\leq 0.
\end{equation}
It follows that the magnetic field inside the \textquotedblleft
bubble\textquotedblright\ is decreasing exponentially from its boundary to
the axis, while the boundary itself is quite \textquotedblleft
rigid\textquotedblright , i.e., there is no pressure at all beyond $r=a$.

Now we can calculate the flux of particles that are lost from the
\textquotedblleft developed bubble\textquotedblright\ in a stationary state.
It is obviously equal to the total source of particles that is necessary to
keep the \textquotedblleft bubble\textquotedblright\ stationary. It is given
in the normalized form by Eq.(\ref{fb}), and can be reconstructed as
\[
\Phi =-\frac{c^{2}L}{4\pi \sigma }\frac{\pi }{2}n_{B}\frac{\beta \beta
^{\prime }r}{1-\beta }=\frac{4}{\sqrt{15}}\frac{a}{\lambda }\frac{c^{2}L}{
4\pi \sigma }\frac{\pi }{2}n_{B}.
\]
Here $n_{B}=B_{v}^{2}/16\pi kT$ is the central density of ions that
corresponds to $\beta =1,$ and 
\[
\frac{c^{2}L}{4\pi \sigma }=4\lambda ^{2}v_{m}\frac{B_{v}}{B_{m}}=\frac{
4\lambda ^{2}v_{m}}{R_{m}}.
\]
It follow that
\begin{equation}
\Phi =\frac{8\pi }{\sqrt{15}}\frac{a\lambda v_{m}}{R_{m}}n_{B}.
\end{equation}

This flux in turn can be used to estimate the particle confinement time: 
\begin{equation}
\tau _{n}=\frac{\pi a^{2}Ln_{B}}{\Phi }\approx \frac{a}{\lambda }\frac{LR_{m}
}{2v_{m}}=\frac{a}{\lambda }\tau _{GDT},
\end{equation}
\begin{equation}
\tau _{n}=\frac{a}{\sqrt{\frac{c^{2}}{4\pi \sigma }\frac{LR}{4v_{m}}}}\frac{
LR_{m}}{2v_{m}}=\sqrt{2\tau _{\perp }\tau _{GDT}},
\end{equation}
where $\tau _{GDT}=LR_{m}/2v_{m}$ is the gas-dynamic time in the vacuum
field, and $\tau _{\perp }=4\pi \sigma a^{2}/c^{2}$ is the transverse
diffusion time over the full \textquotedblleft bubble\textquotedblright\
radius.

\subsection{Analysis}

\begin{enumerate}
\item The \textquotedblleft bubble\textquotedblright\ radius $a$ is directly
proportional and responds to the particle source $\Phi ,$ while the relevant
values and profiles of $\beta $ are then defined selfconsistently. This
means that the equilibrium is quite robust and stable vs. source variations,
although $\beta $ may be exponentially close to the equilibrium limit.

\item The plasma confinement within the \textquotedblleft
bubble\textquotedblright\ can be approximately described as follows. Deep
within it the radial diffusion dominates, while the axial loss is
vanishingly small. In fact the ions may not be magnetized inside of the
\textquotedblleft developed bubble\textquotedblright\ at all, with almost
straight trajectories. All of the radial confinement is concentrated in the
relatively thin boundary layer of width $\sim 6\lambda .$ However,
due to finite magnetic field within the boundary layer, the effective mirror
ratio is also finite, so that the gas-dynamic losses appear. In a unit of
time the \textquotedblleft bubble\textquotedblright\ looses particles from
the layer of width $\lambda $ and radius $a$ via the standard gas-dynamic
axial outflow, hence $\tau _{n}=\frac{a}{\lambda }\tau _{GDT}$.

\item A rough estimate for the confinement quality at reactor parameters ($
T=9keV,$ $L=30m,$ $B_{v}=10T,$ $R_{m}=2,$ $a=1m$) can be obtained using the classical (Spitzer) transverse plasma conductivity:
$
\lambda \approx 2\times 10^{-4}m,
$
$
\tau _{GDT}\approx 5\times 10^{-5}s,
$
and then
\[
\tau _{E}=\frac{3}{8}\frac{a}{\lambda }\tau _{GDT}\approx 0.1s.
\]
In the last estimate we also assumed the GDT scaling of $\tau _{E}\approx 3\tau _{n}/8$ (the plasma energy is proportional to $3T$ per ion, while the axial losses scale as $8T$ per ion.) At $10$ Tesla the central density is $
n_{B}=1.4\times 10^{22}m^{-3};$ as a result
\begin{equation}
n\tau _{E}\approx 1.4\times 10^{21}>10^{20}m^{-3}s.
\end{equation}
Our estimate exceeds the Lawson criterion by a factor of 10. This means that even if the effective plasma resistivity (or radial
diffusion) is a factor of 100 higher than the classical one, the 30m-by-1m DT reactor is still possible.
\item Results of this section are obtained in the MHD approximation, which is not quite applicable. Indeed, the trajectories of ions in reality may be extended far beyond the predicted thin MHD-boundary. Even ions passing right through the middle of the ``bubble'' will reflect back only after passing one Larmor radius into the magnetic field of the border, making its width of the order of $\rho_i\sim c/\omega_{pi}$ (if it is larger than a few $\lambda$). However, this will not necessarily lead to significant increase in the particle losses. In fact, the confinement quality may improve. Indeed, the ions passing both through the border ($B\sim B_v$) and the interior ($B\sim 0$) have very large values of the magnetic moment, i.e., are far from the loss cone, and, from the viewpoint of axial losses in the boundary, are confined in the kinetic rather than in the gas-dynamic regime. This means that their axial losses should be far below the gas-dynamic estimate. If we consider that only ions that do not pass through the ``bubble'' body can be lost, then their loss rate is again limited by radial diffusion into the boundary. However, the rate of this diffusion may be slower than in the MHD case, since the radial gradient of the magnetic field is lower as $\lambda/\rho_i$. On the other hand, the kinetic regime of confinement in linear traps is often unstable, which may increase the loss rate again. The qualitative and quantitative descriptions of kinetic processes within the ``bubble'' boundary are definitely very important and should be addressed in the near future.
\end{enumerate}

\section{\label{sec:St}APPROACHES TO MHD STABILIZATION}
As already noted in the previous sections, the plasma equilibrium in the $\beta\rightarrow 1$ limit is going to be unstable if no special measures are taken. Stabilization will be very tricky and difficult, but, as shown above, potential benefits of using the diamagnetic ``bubble'' confinement are great and can justify any effort. At present the description of the equilibrium is not yet sufficiently advanced and detailed to warrant an in-depth theoretical study of its stability. However, some general ideas of how a ``bubble'' should be formed and confined in order to avoid the most dangerous modes can be discussed. Another possibility is to look for stabilization as in existing analogs, such as FRCs. The outer magnetic configuration of a linear trap in the ``bubble'' regime is equivalent to the FRC scrape-off layer, though the FRC itself is replaced by the low-field ``bubble''. This means that the question of MHD stability of the boundary of the ``bubble'' configuration is mostly similar to that of the FRC, while the inside of the ``bubble'' has far less free magnetic energy and thus should be more stable. We know that the conducting-shell stabilization works in the case of C-2U\cite{C2U} and can look in this direction as well.

Two general recipes for formation of ``bubbles'' are already formulated: 1) the plasma heating scheme should avoid any strong anisotropy of the resulting plasma pressure, and 2) there should be a long patch of nearly-uniform field at the bottom of the magnetic well. The first requirement will allow stability of anisotropy-related modes, while the second one will make the form of the ``bubble'' equilibrium quasi-cylindrical. In the following we assume that these requirements are satisfied.

A lot is already known about the inherent instability of flute modes in linear axisymmetric traps and the various ways to stabilize them.\cite{Ryutovstab} Most of stabilization methods rely on special systems or cells placed at the ends of the traps, or in expanders. As shown, for example, by Ryutov and Stupakov,\cite{Ball1} all of them fail when $\beta$ is large, i.e., when the thermal energy becomes comparable to the magnetic field energy and the field lines can no longer be considered ``rigid''. If the trap is long, the instability-driving curvature is small, but the perturbation of the magnetic field energy due to bending of field lines between the plasma body and the far-away stabilizers is equally small.
Let's apply this qualitative reasoning to our proposed equilibrium. 

The instability-driving curvature at ends of the ``bubble'' cylinder may be significant, but it is localized. Its location is also well-defined, so that external stabilizers can be (and should be) placed nearby (see Fig.\ref{puzyr}). There is also a particular type of such stabilizers, the massive shell conductors for line-tying, that is especially suitable for the purpose. Indeed, the own magnetic field of the quasi-cylindrical plasma column is similar to that of a solenoid, i.e., the own field at the cylinder ends is rapidly expanding out of the plasma. Most of this flux can be easily intercepted by conductors. The distance between conductors and the plasma edge along the own plasma field will be quite small. Also note that the own plasma field at $\beta\sim 1$ is as great as the confining field, so that its bending energy over short distance can be sufficiently large to inhibit ballooning perturbations of the plasma edge. 

Okay, this variant of the conducting-shell stabilizer may be sufficient for suppressing long-wave instabilities of the edge. What about the short waves and the interior? For suppression of short-wave modes we have to rely on the finite-larmor-radius (FLR) effects. They are shown to work quite well in the GDT-like traps\cite{Ivanov, Ryutovstab}. However, because of sharp field gradients the standard perturbative methods fail, and there is no adequate kinetic description of the equilibrium, so that the full theory of FLR stabilization of the ``bubble'' edge should be left for future studies. Still, it is worth to point out the question of balance: if the FLR is large, we cannot place the conducting shell closer than $\rho_i$ to the plasma surface. Fortunately, at least the interior of the ``bubble'' needs no special efforts for stabilization: there are no gradients and no magnetic energy, and thus no instability, since the radial transport in the unmagnetized plasma is already as large as it can be.

\section{\label{sec:Conc}CONCLUSION}
A new scheme for confining high-$\beta$ fusion plasmas in a linear trap is described. It is midway between the classical schemes of the gas-dynamic trap and the FRC, and is not really much different from what was attempted before. However, the present estimates show that the diamagnetic ``bubble'' equilibrium promises huge improvement of confinement quality as compared to the gas-dynamic scheme, so that the estimated reactor length reduces from 5km to 30m and the total fusion power becomes reasonable. The confinement time scales as the geometric average of the gas-dynamic time and the time of the radial diffusion of the magnetic flux, $\tau_E\approx\sqrt{\tau_\parallel\tau_\perp}$. A stable confinement of the $\beta\approx 1$ plasma cannot be easy, but there seems to be a straightforward way to use the conducting-shell stabilization method that is shown to work for FRCs.\cite{C2U}

Although there is still no detailed theory of stability and transport, it is probably worthwhile to attempt an experimental check of the predicted ``bubble'' formation and of the related improvement in confinement time. Such initial concept-exploration experiments are now in the planning stage in the Budker Institute of Nuclear Physics in Novosibirsk. The common problems are the requirement of a very high power density of heating during the formation stage (as detailed in section III of this paper), and the necessity to have a programmable particle source within the ``bubble'' (that can be achieved by pellet-injection type systems in larger traps). One approach is to use highly-focused nearly-transverse NBI heating in the small CAT mirror (in design stage), which is originally intended for an attempt to create a beam-driven FRC. The second possible approach is to use RMF heating like in the PFRC project in the Princeton PPL.\cite{Princeton} 

Experiments to form FRCs are generally very similar to those to form a ``bubble''. The main difference lies in the plasma conductivity in the formation stage. While during the FRC formation the magnetic flux in its interior has to reconnect, which would be difficult in highly-conducting plasma, in the ``bubble'' case there is no reconnection. Instead, the flux-conserving radial expansion of flux tubes can lead to a drastic drop of the ion density. Without solving the problem of feeding the ``bubble'' by sufficient numbers of particles, the experiments would produce transient states at best. The simple and common way of the gas puff is clearly not the best choice: it cannot be applied in the area of fast-ion confinement for fear of charge-exchange losses, while outside of this area the particle feed will not reach the core of the ``bubble''. However, the author is sure that the ingenuity of physicists is limitless and the still outstanding problems can be solved one way or another.

\begin{acknowledgments}
The author is grateful to his colleagues from the Budker Institute for encouragement and many fruitful discussions, especially to V.V. Postupaev for commenting the manuscript; and
to Prof. T. Tajima of Tri Alfa Energy, Inc. for suggesting an important reference.

This work has been supported by Russian Science Foundation (project N 14-50-00080).
\end{acknowledgments}


%
%



\end{document}